\documentclass[runningheads]{llncs}
\usepackage[T1]{fontenc}
\usepackage{graphicx}
\usepackage[frozencache]{minted}
\usepackage{hyperref}
\usepackage{color}

\begin{document}
\title{Python client for Isabelle server\thanks{This work has been supported by the French government, through the 3IA Côte d'Azur Investments in the Future project managed by the National Research Agency (ANR) with the reference number ANR-19-P3IA-0002.}}
%
%
\author{Boris Shminke\inst{1}\orcidID{0000-0002-1291-9896}}
\authorrunning{B. Shminke}
\institute{Université Côte d'Azur, CNRS, LJAD, France \\
\email{boris.shminke@univ-cotedazur.fr}}
\maketitle
                     
\begin{abstract}
We contribute a Python client for the Isabelle server, which gives researchers and students using Python as their primary programming language an opportunity to communicate with the Isabelle server through TCP directly from a Python script. Such an approach helps avoid the complexities of integrating the existing Python script with languages used for Isabelle development (ML and Scala). We also describe new features that appeared since the announcement of the first version of the client a year ago. Finally, we give examples of the client's applications in research and education and discuss known limitations and possible directions for future development.
                     
\end{abstract}
\section{Introduction}
Isabelle~\cite{DBLP:books/sp/NipkowPW02} interactive theorem prover (ITP) has included the Isabelle server as part of its standard distribution since 2018~\cite{IsabelleServer}. The Isabelle server enables users to run multiple sessions and manage concurrent tasks to process Isabelle theory files through TCP. It makes, in principle, possible to communicate with the Isabelle server using any popular programming language~\cite{TIOBEIndex}, including Python. Python clients already exist for other major ITPs, for example, one~\cite{LeanClient} for Lean~\cite{10.1007/978-3-030-79876-5_37} or another one~\cite{CoqClient} for Coq~\cite{the_coq_development_team_2022_5846982}. To our best knowledge, the previous version (0.2.0) of the \texttt{isabelle-client} announced last year~\cite{DBLP:conf/mkm/LiskaLNRSSSW21} was the first Python client for Isabelle. This paper describes version 0.3.5 of the client (the digital artefact is available on Zenodo~\cite{boris_shminke_2022_6490275}).

\section{How to use}
\begin{figure}
\includegraphics[width=\textwidth]{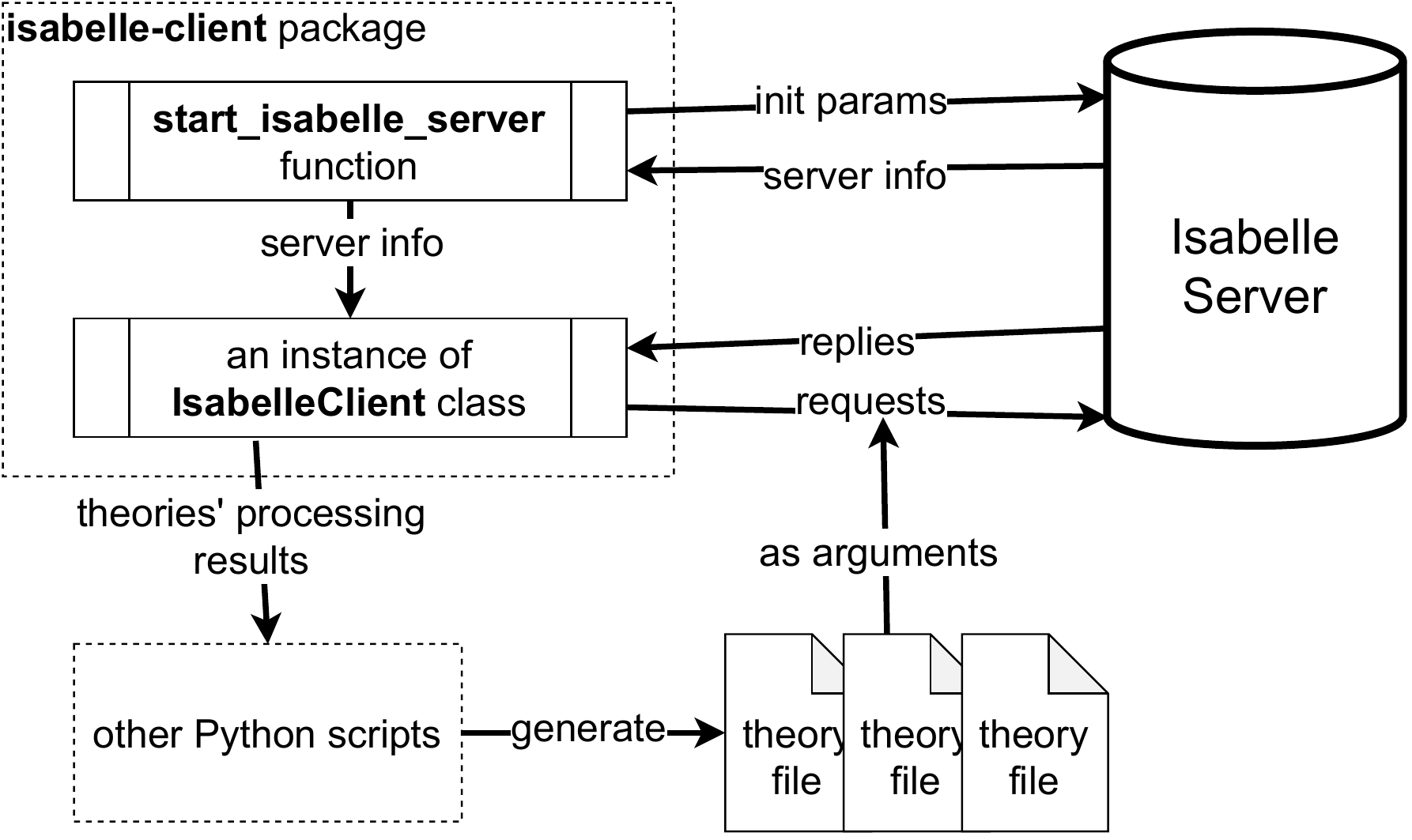}
\caption{In a project where Python scripts are generating Isabelle theory files, \texttt{isabelle-client} can be used to start the Isabelle server, sending these files for parallel processing and getting back the results.} \label{fig:architecture}
\end{figure}

A typical scenario of \texttt{isabelle-client} application is represented by Fig.~\ref{fig:architecture}. First, one should start the Isabelle server, for example, using a utility function \texttt{start\_isabelle\_server} from the package. Second, one creates an instance of \texttt{IsabelleClient} object, e.g. using the factory function \texttt{get\_isabelle\_client}. Finally, one can issue any command supported by the Isabelle server (see section 4.4 in~\cite{IsabelleSystemManual} for a full list) using \texttt{IsabelleClient} object, which implements all these commands as methods. Usually, these commands rely on existing Isabelle theory files, for example, generated by third-party Python scripts (not by \texttt{isabelle-client}). See also listing \ref{lst:example} for a basic code snippet.

\begin{listing}[H]
\begin{minted}{python}
""" An example of the ``isabelle-client`` usage """
from isabelle_client import get_isabelle_client, start_isabelle_server

# first, we start Isabelle server
server_info, _ = start_isabelle_server(
    name="test", port=9999, log_file="server.log"
)
# then we create an ``IsabelleClient`` instance
isabelle = get_isabelle_client(server_info)
# now we can send theory files to the server and get a response
isabelle.use_theories(theories=["Example"], master_dir=".")
# or we can build a session document using ROOT and root.tex files
isabelle.session_build(dirs=["."], session="examples")
isabelle.shutdown()
\end{minted}
\caption{How to use \texttt{isabelle-client}.}
\label{lst:example}
\end{listing}

Interested potential users of the client can also follow the package homepage~\footnote{\href{https://pypi.org/project/isabelle-client/}{https://pypi.org/project/isabelle-client/}} for more detailed information.

\section{New features since previous version}
The first public version of this client~\cite{DBLP:conf/mkm/LiskaLNRSSSW21}, which appeared in March 2021, worked only for Python 3.7 on GNU/Linux and was supposed to be installed only with the \texttt{pip} package manager. The current version is available for any Python 3.6+ on GNU/Linux and Windows. Every new build is tested in a continuous integration workflow against each supported Python version. The package is hosted now not only on the Python Package Index (PyPI), but also on Conda Forge~\cite{conda_forge_community_2015_4774216}, which enables its installation with both \texttt{pip} and \texttt{conda} package managers. In addition, one can run the client inside a Docker container, for example, in a cloud using Binder~\cite{project_jupyter-proc-scipy-2018}. This option provides the client coupled with the Isabelle server and is particularly useful for students specialising in logic but not necessarily having much experience in information technologies.

The current version of the client is tested to work with the latest Isabelle 2021-1 (released in December 2021). The last client version returns all server replies as a Python list, not only the last one as it was in the previous version giving more flexibility to the end-user. Last but not least, the current version arrives with detailed documentation pages and is nearly 100\% covered with unit tests using fixtures for emulating a working Isabelle server behaviour.

\section{Known applications in research and education}
The original system entry for \texttt{isabelle-client} expressed hope that it would be helpful to other projects. After the first year, we saw several applications confirming these aspirations. The most impressive one was a discovery of a proof~\cite{DBLP:journals/corr/abs-2109-05264} for an algebraic problem which stood open for two years despite the efforts of specialists in the field. For this project, the \texttt{isabelle-client} played a role of a bridge connecting Python scripts which generated relevant Isabelle theory files with the Isabelle server processing them.

Another use case comes from higher education. The \texttt{isabelle-client} running in a Docker container on Binder was used during the practical sessions of the Advanced Logic course taught at the Université Côte d'Azur in the autumn of the 2021-2022 academic year. The client helped students not trained in functional programming languages used for Isabelle development to concentrate on understanding the Isabelle language syntax and consequently generating theory files in it with Python scripts without installing and running the Isabelle GUI on their laptops.

Also, Fabian Huch used the \texttt{isabelle-client} for debugging the 'Proving for Fun' backend~\cite{haslbeck2019competitive}. The 'Proving for Fun' project already had in its codebase parts resembling the \texttt{isabelle-client}, but using the \texttt{socket} Python package that provides access to the BSD socket interface. The \texttt{isabelle-client} uses \texttt{asyncio}, providing a more high-level interface for TCP communications, and exists as a stand-alone piece of software striving to be easily reusable for different projects.

\section{Known limitations, workarounds, and future work}
The first thing one would want from a client-server application is running the server and the client on different machines (probably, with various operating systems) or, at least, in separate Docker containers. The attempts to do that with \texttt{isabelle-client} demonstrated that the current implementation of the Isabelle server doesn't support binding its IP address to anything but \texttt{localhost} for security reasons. A desirable solution is implementing an authentication mechanism for the remote Isabelle server (for example, with access tokens, as is done for the Jupyter~\cite{DBLP:journals/cse/GrangerP21} server). The current workaround is to use SSH tunnels or package the \texttt{isabelle-client} together with the Isabelle server in the same Docker container.

Although the \texttt{isabelle-client} works perfectly on Windows for communication with the Isabelle server running on the same machine, it's impossible to start the Isabelle server on Windows with the utility function from the \texttt{isabelle-client} package. The reason is related to the fact that the Isabelle server on Windows runs under Cygwin and not as a proper Windows application. A possible solution could be to rewrite the Windows CMD file, which starts the Isabelle server, to make it pass the started server info from the Cygwin subsystem to the parent Windows process. The current workaround is to launch the Isabelle server manually.

At some point, it became known that it's impossible to start a clean build of a session from the \texttt{isabelle-client}. The reason is that the Isabelle server (unlike \texttt{isabelle build} tool) lacks \texttt{-c} parameter in its API. Thus, achieving relative feature parity of the Isabelle server and other Isabelle tools seems to be an important direction in making the \texttt{isabelle-client} more applicable and interesting to its end users.
\section{Conclusion}
A year after its appearance \texttt{isabelle-client} became a more mature piece of software and found its applications in both scientific research and higher education. As shown in this paper, there are still many ways to improve it, and we hope for many possible occasions for its further application, not limited to but including machine learning (as a domain dominated by Python~\cite{KaggleReport}) for theorem proving.
\bibliographystyle{splncs04}
\bibliography{cicm2022}

\begin{thebibliography}{10}
\providecommand{\url}[1]{\texttt{#1}}
\providecommand{\urlprefix}{URL }
\providecommand{\doi}[1]{https://doi.org/#1}

\bibitem{conda_forge_community_2015_4774216}
conda-forge community: {The conda-forge Project: Community-based Software
  Distribution Built on the conda Package Format and Ecosystem} (Jul 2015).
  \doi{10.5281/zenodo.4774216}, \url{https://doi.org/10.5281/zenodo.4774216}

\bibitem{DBLP:journals/corr/abs-2109-05264}
Fussner, W., Shminke, B.: Mining counterexamples for wide-signature algebras
  with an isabelle server. CoRR  \textbf{abs/2109.05264} (2021),
  \url{https://arxiv.org/abs/2109.05264}

\bibitem{DBLP:journals/cse/GrangerP21}
Granger, B.E., P{\'{e}}rez, F.: Jupyter: Thinking and storytelling with code
  and data. Comput. Sci. Eng.  \textbf{23}(2),  7--14 (2021).
  \doi{10.1109/MCSE.2021.3059263},
  \url{https://doi.org/10.1109/MCSE.2021.3059263}

\bibitem{haslbeck2019competitive}
Haslbeck, M.P., Wimmer, S.: Competitive proving for fun. Kalpa Publications in
  Computing  \textbf{10},  9--14 (2019)

\bibitem{project_jupyter-proc-scipy-2018}
{P}roject {J}upyter, {M}atthias {B}ussonnier, {J}essica {F}orde, {J}eremy
  {F}reeman, {B}rian {G}ranger, {T}im {H}ead, {C}hris {H}oldgraf, {K}yle
  {K}elley, {G}ladys {N}alvarte, {A}ndrew {O}sheroff, {P}acer, M., {Y}uvi
  {P}anda, {F}ernando {P}erez, {B}enjamin~{R}agan {K}elley, {C}arol {W}illing:
  {B}inder 2.0 - {R}eproducible, interactive, sharable environments for science
  at scale. In: {F}atih {A}kici, {D}avid {L}ippa, {D}illon {N}iederhut,
  {P}acer, M. (eds.) {P}roceedings of the 17th {P}ython in {S}cience
  {C}onference. pp. 113 -- 120 (2018). \doi{10.25080/Majora-4af1f417-011}

\bibitem{DBLP:conf/mkm/LiskaLNRSSSW21}
L{\'{\i}}ska, M., Lupt{\'{a}}k, D., Novotn{\'{y}}, V., Ruzicka, M., Shminke,
  B., Sojka, P., Stef{\'{a}}nik, M., Wenzel, M.: Cicm'21 systems entries. In:
  Kamareddine, F., Coen, C.S. (eds.) Intelligent Computer Mathematics - 14th
  International Conference, {CICM} 2021, Timisoara, Romania, July 26-31, 2021,
  Proceedings. Lecture Notes in Computer Science, vol. 12833, pp. 245--248.
  Springer (2021). \doi{10.1007/978-3-030-81097-9\_20},
  \url{https://doi.org/10.1007/978-3-030-81097-9\_20}

\bibitem{10.1007/978-3-030-79876-5_37}
Moura, L.d., Ullrich, S.: The lean 4 theorem prover and programming language.
  In: Platzer, A., Sutcliffe, G. (eds.) Automated Deduction -- CADE 28. pp.
  625--635. Springer International Publishing, Cham (2021)

\bibitem{DBLP:books/sp/NipkowPW02}
Nipkow, T., Paulson, L.C., Wenzel, M.: Isabelle/HOL - {A} Proof Assistant for
  Higher-Order Logic, Lecture Notes in Computer Science, vol.~2283. Springer
  (2002). \doi{10.1007/3-540-45949-9},
  \url{https://doi.org/10.1007/3-540-45949-9}

\bibitem{CoqClient}
\relax{Gallego Arias}, E.J., Martinez, T.: Pycoq: Access coq from python! (Jan
  2022), \url{https://github.com/ejgallego/pycoq}

\bibitem{KaggleReport}
\relax{Kaggle Inc.}: State of data science and machine learning 2021 (Oct
  2021), \url{https://www.kaggle.com/kaggle-survey-2021}

\bibitem{the_coq_development_team_2022_5846982}
\relax{The Coq Development Team}: The coq proof assistant (Jan 2022).
  \doi{10.5281/zenodo.5846982}, \url{https://doi.org/10.5281/zenodo.5846982}

\bibitem{TIOBEIndex}
\relax{TIOBE Software BV}: The tiobe programming community index (Apr 2022),
  \url{https://www.tiobe.com/tiobe-index/}

\bibitem{LeanClient}
Rute, J., Massot, P., Berman, J., Roux, F.L.: Lean client for python (Aug
  2021), \url{https://github.com/leanprover-community/lean-client-python}

\bibitem{boris_shminke_2022_6490275}
Shminke, B.: Python client for isabelle server (Apr 2022).
  \doi{10.5281/zenodo.6490275}, \url{https://doi.org/10.5281/zenodo.6490275}

\bibitem{IsabelleServer}
Wenzel, M.: Isabelle/pide after 10 years of development. In: 13th International
  Workshop on User Interfaces for Theorem Provers (UITP 2018). Federated Logic
  Conference 2018 (2018),
  \url{https://sketis.net/wp-content/uploads/2018/08/isabelle-pide-uitp2018.pdf}

\bibitem{IsabelleSystemManual}
Wenzel, M.: The isabelle system manual (Dec 2021),
  \url{https://isabelle.in.tum.de/dist/Isabelle2021-1/doc/system.pdf}

\end{thebibliography}
\end{document}